\documentclass[preprint,11pt,preprintnumbers,nofootinbib]{revtex4}
\usepackage{graphicx}
\usepackage{dcolumn}
\usepackage{bm}
\usepackage{amsfonts}

\begin{document}
\date{\today}
\title{Exposition to Staggered Multi-Field Inflation}

\author{Thorsten Battefeld}
\email{tbattefe(AT)princeton.edu}
\affiliation{ Princeton University,
Department of Physics,
NJ 08544
}

\begin{abstract}
I investigate multi-field inflationary models with fields that decay during inflation, leading to staggered inflation. This feature is natural in many models motivated by string theory, for instance if inflatons are related to interbrane distances and the branes start to annihilate during inflation. A short exposition to an analytic framework is provided, enabling the computation of leading order corrections to observables, i.e. the scalar spectral index.
\end{abstract}
\maketitle

\section{Framework}
First, I give a short introduction to staggered multi-field inflation \cite{Battefeld:2008py}. Consider, for simplicity, $\mathcal{N}$ uncoupled scalar fields $\varphi_A$, $A=1\dots \mathcal{N}$, starting out from identical field values and with identical potentials $V_A\equiv V$, such that assisted inflation \cite{Liddle:1998jc} (slow roll inflation driven by many fields) results. If fields start to decay during inflation, for instance because they are related to branes that annihilate \cite{Becker:2005sg} or tachyons that condense \cite{Majumdar:2003kd}, the number of fields decreases, causing a decrease of the energy driving inflation $\rho_{\mathcal{I}}\simeq \mathcal{N}V\equiv W$. The energy in the just decayed field does not vanish, but gets converted into some other form $\rho_r$, i.e. radiation. I take $p_r=w_r\rho_r$ with $w_r=const$ and set $8\pi G\equiv 1$ from here on.

To recover this effect within a simple analytic framework, I average the number of fields over an interval longer than the decay time, but shorter than the Hubble time, such that $\mathcal{N}(t)$ becomes a smooth function. Due to this averaging any effects caused by the sudden drops in $\rho_{\mathcal{I}}$, such as a ringing in the power-spectrum, cannot be recovered. Hence, this is only a good approximation if several fields decay during any given Hubble time. Given this smoothing, one can introduce a decay rate $\Gamma \equiv -\dot{\mathcal{N}}/\mathcal{N}>0$, which is a prescribed, albeit model specific function. In order to prevent a premature end of inflation we need 
\begin{eqnarray}
\bar{\varepsilon}&\equiv& \frac{3}{2}(1+w_r)\frac{\rho_r}{\rho_{\mathcal{I}}+\rho_r}\ll 1 \,, \label{defbarepsilon}\\
\varepsilon_{\mathcal{N}}&\equiv& \frac{\Gamma}{2H}\ll 1 \,. \label{defepsilonn}
\end{eqnarray}
The background dynamics, including the energy transfer from $\rho_{\mathcal{I}}$ to $\rho_{r}$, is then given by the Friedman equations and the continuity equations 
\begin{eqnarray}
\dot{\rho}_{\mathcal{I}} \!\! &=& \!\! -3H(\rho_{\mathcal{I}}+p_{\mathcal{I}})+\dot{\mathcal{N}}V\simeq -2H(\varepsilon_{\mathcal{N}}+\varepsilon)\rho_\mathcal{I}\,,\\
\dot{\rho}_r\!\! &=& \!\! -3H(\rho_r+p_r)-\dot{\mathcal{N}}V\simeq2H (\varepsilon_{\mathcal{N}}-\bar{\varepsilon})\rho_{\mathcal{I}} \,,
\end{eqnarray}
so that $\nabla_\mu T^{\mu 0}_{\mbox{\tiny total}}=0$. Here, $\varepsilon \equiv (W^\prime/W)^2/2\ll 1$ is the slow roll parameter of the effective single field $\varphi\equiv \sqrt{\mathcal{N}}\varphi_A$, $W^\prime=\partial W/\partial \varphi$ and the ``$\simeq$'' denotes equality to first order in small parameters. This framework is reminiscent of warm inflation \cite{Berera:1995ie}, with the notable difference that potential energy is converted to $\rho_r$ as opposed to kinetic, thus avoiding many of the problems associated with warm inflation.

During inflation, the Hubble slow roll parameter becomes $-\dot{H}/ H^2\simeq \varepsilon + \bar{\varepsilon}$ and one can show that $\rho_r$ approaches a scaling solution $\rho_r\rightarrow  \varepsilon_{\mathcal{N}} 2\rho_{\mathcal{I}}/(3+3w_r)$ with $\dot{\rho}_r\sim \mathcal{O}(\varepsilon_{\mathcal{N}})\rho_r$ for which $\varepsilon_{\mathcal{N}}\simeq \bar{\varepsilon}$  \cite{Battefeld:2008py}.

\section{Perturbations} 
Due to the presence of many scalar fields and $\rho_r$, isocurvature perturbations could be produced \cite{Gordon:2000hv}. However, since all the scalar fields are rolling slowly and $\rho_r$ scales like $\rho_{\mathcal{I}}$, the nonadiabatic pressure \cite{Gordon:2000hv} is slow roll suppressed \cite{Battefeld:2008py} (see also \cite{Watson:2006px}). Therefore, one can focus on adiabatic perturbation, which are correctly recovered by employing  the Mukhanov variable $v_k$ \cite{Mukhanov:1990me}. Following standard techniques \cite{Mukhanov:1990me}, one can then compute the power spectrum $\mathcal{P}_\zeta=k^3|\zeta_k|^2/(2\pi^2)$ of the curvature perturbation on uniform density hyper-surfaces. Using $\bar{\varepsilon}\simeq \varepsilon_{\mathcal{N}}$, the scalar spectral index $n_s-1=\partial \ln \mathcal{P}_\zeta / \partial \ln k$ becomes \cite{Battefeld:2008py}
\begin{eqnarray}
\nonumber n_s-1\!&\simeq&\!-2(\varepsilon+\bar{\varepsilon})-\frac{2}{\varepsilon\gamma^2+\bar{\varepsilon}} \bigg[\varepsilon\gamma^2(2\varepsilon-\bar{\varepsilon}-\eta)\\
&&+(\varepsilon+\bar{\varepsilon})(1-\delta)(\varepsilon\gamma(\gamma-1)+\frac{\bar{\varepsilon}}{2})\bigg],\label{finalns}
\end{eqnarray}
where  $\eta\equiv W^{\prime\prime}/(2W)$ and $\gamma\equiv 1+\varepsilon_{\mathcal{N}}\varphi \frac{W}{W^\prime}$ as well as $\delta\equiv \dot{\Gamma}H/(\Gamma \dot{H})$. One recovers the usual slow roll result for $\Gamma=0$, that is in the limit $(\bar{\varepsilon},\varepsilon_{\mathcal{N}})\rightarrow 0$. On the other hand,  $(\varepsilon,\eta, \varepsilon\gamma^2) \rightarrow 0$ while $\varepsilon_{\mathcal{N}}=const$ so that $\delta=1$, corresponds to a dynamically relaxing cosmological constant with $n_s-1\simeq -2\bar{\varepsilon}$, which is discussed in great detail in \cite{Watson:2006px}. Similarly, one can compute the running, which remains second order in small parameters and is thus unobservable small. Other observables such as non-Gaussianities or tensor perturbations could also be computed \cite{preparation}.

\section{Application}
For instructive purposes I would like to apply the formalism to the model of \cite{Majumdar:2003kd}, where inflation is driven by $\mathcal{N}\sim\mathcal{O}(10^3)$ uncoupled tachyons (related to D-brane/antibrane pairs) that condense during inflation in a staggered fashion. Each tachyon sits close to the top of a hilltop-potential \cite{Majumdar:2003kd} and whenever a quantum mechanical fluctuation displaces a field, it condenses rapidly. If the fields are very close to the maximum, the slow roll contributions become negligible and one can use $n_s-1\simeq \bar{\varepsilon}(\delta-3)$. Note that the amplitude of perturbations (set by the COBE normalization) is determined by the decay rate, that is by $\varepsilon_{\mathcal{N}}$ (just as in \cite{Watson:2006px}), and not by $\varepsilon$. Considering a constant decay rate with $\delta=0$ one can compute the number of e-folds before $\varepsilon_{\mathcal{N}}\sim \mathcal{O}(1)$ to $N \approx  2\Gamma^{-1}\left(\tau_p\mathcal{N}_0/3\right)^{1/2}$ where $\mathcal{N}_0$ is the initial number of fields and $\tau_p$ a model specific brane tension of order one. To achieve sixty e-folds of inflation one needs a few hundred fields and a decay rate slightly smaller than one. Further, $\bar{\varepsilon}\simeq \varepsilon_{\mathcal{N}}\approx 1/N$, so that $n_s-1\approx -3/N$. With $N\sim 60$, this is well within the $1\sigma$ error bars of WMAP5 \cite{Komatsu:2008hk}. Of course one can investigate other decay rates, incorporate slow roll contributions and consider different models as well \cite{Battefeld:2008py}. 

\section{Outlook}
I investigated an analytic formalism and some consequences of staggered multi-field inflation, which arises when inflaton-fields decay during inflation. In future studies one could extend the formalism by relaxing simplifying assumptions, compute other observables such as non-Gaussianities and gravitational waves \cite{preparation}, or construct new models based on the staggered inflation effect.

\begin{acknowledgments}
I am thankful to my collaborators D.~Battefeld and A.~C.~Davis, as well as P.~Steinhardt, S.~Watson and N.~Ashorion for comments. T.~B.~is supported by the Council on Science and Technology at Princeton, and acknowledges hospitality at the Institut d'etudes Sciencifiques de Cargese.
\end{acknowledgments}

\end{document}